# Correlated observables in single-particle systems and field theoretic interpretations

Ian T. Durham[†]

[†]*Department of Physics, Saint Anselm College,*
*100 Saint Anselm Drive, Box 1759, Manchester, NH 03102, USA*
*(idurham@anselm.edu)*
(Updated: November 25, 2005)

**Abstract**

Bell-type experiments that test correlated observables typically involve measurements of spin or polarization on multi-particle systems in singlet states. These observables are all non-commuting and satisfy an uncertainty relation. Theoretically, the non-commuting nature should be independent of whether the singlet state consists of multiple particles or a single particle. Recent experiments in single neutron interferometry have in fact demonstrated this. In addition, if Bell-type inequalities can be found for experiments involving spin and polarization, the same should be true for experiments involving other non-commuting observables such as position and momentum as in the original EPR paper. As such, an experiment is proposed to measure (quantum mechanically) position and momentum for a single oscillator as a means for deriving a Bell-type inequality for these correlated observables. The experiment, if realizable, would shed light on the basic nature of matter, perhaps pointing to some form of self-entanglement, and would also help to further elucidate a possible mechanism behind the Heisenberg uncertainty principle. Violation of these inequalities would, in fact, offer yet another confirmation of the principle.

## I. INTRODUCTION

Bell-type experiments in their original incarnation were designed to test the correlation of spin in entangled states [1]. The entanglement problem is a long-standing one in quantum mechanics and though many physicists are content to accept its existence without explanation, further elucidation of the problem is required for advances in quantum computing, quantum cryptography, and some areas of nanotechnology. In the typical Bell-type situation a pair of entangled electrons are emitted from a common source in opposite directions and their spins are measured along a given axis. The correlation of the spins due to the entangled nature of the electrons (which are in a singlet state), along with the Pauli exclusion principle, requires that if one electron should be forced to change its spin, the other electron must do so as well *instantaneously*. Seemingly, there is a violation of the special relativistic speed-of-light limit on the transmission of information.

Early on, realistic experiments focused on the easier task of modeling the Bell-type situation with entangled photons where the polarization angles of the photons took the place of the projections of the spin of the electrons onto a specific axis [2]. Clearly one of the most unusual properties of entanglement that was demonstrated in these experiments was the seemingly superluminal signaling that was taking place between the entangled particles. In order to better understand the superluminal signaling, however, it is necessary to better understand the nature of the actual entanglement.

Mathematically the entanglement is represented by non-commuting observables that are experimentally correlated. The wavefunction for entangled systems represents a singlet state and the outcome of the experimental measurements can be assigned probabilities. The probabilities, or a related correlation function that involves the probabilities, typically obey a set of inequalities. Bell's theorem essentially requires that any application of a principle of locality (meaning there can be no superluminal transmission of information) will result in a set of inequalities. These inequalities are violated when apparent superluminal signaling takes place.

There are actually two problems here that bear investigation. The first is the most obvious: the nature of this apparent superluminal signaling. The second is perhaps less obvious but may have a bearing on the first: the nature of the non-commutation itself. In general, Bell-type experiments involve multiple particles. However, the nature of these experiments implies that the nature of the correlation involves the non-commuting observables themselves. As such this implies that a similar type of experiment could be performed on single particles. Indeed entanglement between two degrees of freedom has been demonstrated in single neutron interferometry experiments [3]. Since the entangled observables are always non-commuting they obey an uncertainty relation. In order to explore both the single-particle nature of entanglement and its relation to the uncertainty principle, a Bell-type experiment is proposed to test position and momentum for a single harmonic oscillator.

## II. CONTINUOUS AND DISCRETE MEASUREMENTS

In Bell-type experiments the outcomes of measurements are simplified by the nature of the measuring apparatus. For instance, a particle's quantum mechanical spin is typically measured by passing the particle through an external magnetic field that forces the spin axis to either align or anti-align with the direction of the external field. This is a discrete and essentially binary measurement that normally appears in the operational mathematics as $\pm 1$ representing the spin as either aligned or anti-aligned. A similarly natural binary representation is not obviously evident for position and momentum measurements. For instance, while spin maintains this binary-like feature for molecules, position and momentum appear continuous on a similar scale – they can take on a nearly continuous wide range of values. Data sampling methods such as analog-to-digital conversion, however, can be utilized to overcome this impediment if implemented properly. This is a necessary step for the derivation of a Bell-type inequality since most such inequalities are based on $\pm 1$ (up/down) measurements. A derivation of an equivalent set of inequalities for the correlation of position and momentum for a single particle is dependent, then, upon the experimental and data sampling apparatus.

## III. EXPERIMENTAL APPARATUS

Recent advances in nanotechnology offer device sensitivities approaching the quantum limit making an experiment of this nature closer to reality. The experiment would require simultaneous, independent measurements of position and momentum in order to test the quantum mechanical correlation of these two observables. Laser atom trapping might serve as one potential solution to the "front-end" of such an experiment. Another potential "front-end" solution would be a nano-scale vibrating bar (metallic, organic, etc.) whose motion was monitored by similarly sized nano-scale lasers such as photonic-wire lasers [4] that, for instance, were placed at the extrema of the bar's oscillatory motion. One set of lasers would simply measure the position of the oscillator as a simple triggering mechanism. A second set of lasers would operate in a manner similar to a photogate to measure velocity from which momentum can be easily extracted assuming the mass of the oscillator does not change. A single-electron transistor (SET) might be utilized as the oscillating device [5].

In either case the data sampling equipment measuring position and momentum must be completely independent systems and cannot interfere with each other outside of the actual correlation being studied. Since the independent detectors can theoretically report any value for position or momentum, the system is essentially continuous (note: not in terms of sampling intervals but rather in terms of sampled *values*). Normally a continuous system such as this would employ an analog-to-digital converter (ADC) to extract discrete values from this data. For extremely rapid conversion, a flash ADC is usually employed. In order to approach the quantum limits in sampling, the flash ADC could be constructed utilizing SETs [5-6]. The rapid measurement itself, however, poses a problem for collecting this data in typical memory devices. As such a sample-and-hold amplifier (also constructed with a SET) might need to be inserted in the circuit between the actual measurement device and the flash ADC [7].

The experimental apparatus, in short, would consist of an object – perhaps an atom or a nano-scale bar (modeled potentially by a SET) – in oscillatory motion monitored by two independent sets of lasers (photonic-wire lasers, perhaps, in the case of the oscillating bar) whose output passes through a sample-and-hold amplifier before being processed by a flash ADC where both the amplifier and ADC are constructed with SETs.

## IV. SAMPLING METHODS

The analog-to-digital conversion is made in order to put the results into a binary-like (up/down) form similar to those achieved in typical Bell-type experiments. Tight measurement limits, found by reducing $\Delta x$ and $\Delta p$ to limits normally prevented by the uncertainty principle, can be used as a starting point for achieving the ±1 results. The spread in these measurements can be taken as being over a series of measurements on the given observable (i.e. the spread in the distribution of measured results). Therefore, assign +1 to a measurement (of either $\Delta x$ or $\Delta p$) that exceeds the quantum limits imposed by the uncertainty principle and assign a -1 that does *not* exceed these limits. In short, a +1 is considered to be a *sharp* measurement while -1 is considered to be *fuzzy*.

In order to reduce probability terms in the final inequality it is necessary to reduce, by as much as possible, the probability of error arising from detector failure. As a testing method, then, each detector ($X$ and $P$) should be subjected to repeated, individual, *independent* testing (i.e. testing while the other detector is turned off). Ideally in this situation a perfect detector should measure a +1 every time. Certainly a probability of measuring a -1 in this situation ought to be less than 1% at the bare minimum. This should ensure that any measurement of -1 during joint sampling is entirely due to quantum mechanical effects and not technical problems in the apparatus. Note that this also implies that the probability of a double-negative (a -1 on both detectors simultaneously) ought to be even lower, theoretically, making it almost equivalent to what we would expect with a double-positive. That is, the uncertainty principle should prevent a measurement of +1 on both detectors in a simultaneous measurement meaning the probability of this occurring should be exactly zero. While the probability of a double-negative will not be *exactly* zero, if the probabilities of single-negative measurements are sufficiently small, the double-negative case, being multiplicative, can be arbitrarily pushed close enough to zero as to make it indistinguishable from the double-positive case within the normal sampling uncertainty in the data.

The above argument implies that the highest probabilities and therefore the expected results will be for the mixed cases of +1/-1 and -1/+1, interpreted as a sharp measurement of *either* position or momentum for the joint observation.

## V. BELL-TYPE INEQUALITIES

Following the method recently employed by Andersson, Barnett, and Aspect [8] denote the measurement results as $X_J$, $X'_J$, $P_J$, and $P'_J$ where, once again, the results will all be ±1. Since there is no spatial separation between the sources of these two measurements (it's the same source particle) the superluminal signaling limit imposed by operational locality no longer applies. In a similar vein, however (since the operational locality principle is essentially classical in nature), there is a classical assumption that measurements of these two observables *should* be completely independent of one another. Nonetheless, assume detector $P$ initially returns a given value (denoted simply by $P$ meaning it could be either ±1). The probability that detector $X$ obtains $X_J = X'_J$ (meaning two subsequent measurements by detector $X$ are equal) can be written

$$p(X_J = X'_J) = p(X_J = X'_J = P) + p(X_J = X'_J = -P)$$

where the probabilities on the right side of this equation must exist. This means that they are greater than or equal to zero and thus we can write

$$p(X_J = X'_J = P) + p(X_J = X'_J = -P) \geq |p(X_J = X'_J = P) - p(X_J = X'_J = -P)| \quad (1)$$

The right-hand side of (1), without the absolute value requirement, can be represented by a combination of correlation functions each having the form

$$E(X,P) = p(X = P) - p(X = -P) = \overline{XP}$$

where I have retained Andersson, et. al.'s use of an overbar to represent an average for a given state since, for the derivation, observables and operators have not been defined for the joint measurement at this point. The right side of (1) can thus be written

$$p(X_J = X'_J = P) - p(X_J = X'_J = -P)$$
$$= \frac{1}{2}[E(X_J, P) + E(X'_J, P)]$$

which leads to

$$p(X_J = X'_J) \geq \frac{1}{2}[E(X_J, P) + E(X'_J, P)]. \quad (2)$$

In a similar manner, assume detector $P$ has made just the opposite measurement. A derivation similar to the one just given will yield

$$p(X_J = -X'_J) \geq \frac{1}{2}[E(X_J, P') - E(X'_J, P')]. \quad (3)$$

The probabilities on the left side of (2) and (3) should be independent of the measurement result of *P* since they are, in fact, the sum of all probabilities for a given measurement on *P*. In addition they should add to 1 in order to give the sum of all probabilities for all measurements on *P*. Therefore

$$|E(X_J, P) + E(X'_J, P)| + |E(X_J, P') - E(X'_J, P')| \leq 2. \quad (4)$$

As pointed out by Andersson, et. al., this inequality greatly resembles a CHSH inequality [9]. In addition they note that this inequality *must be satisfied* for joint measurements in quantum mechanics since it is necessary for joint probability distributions to exist for jointly measured observables regardless of whether the measurement is quantum mechanical or classical. Pitowsky has also provided an argument claiming that it is logico-mathematically impossible for any Bell-type inequality to be violated for a *single sample* [9]. All such violations result from *dual* samples. A dual sample amounts to using two detectors – one for each of the two correlated observables. Technically the experiment proposed here is also a dual sample and thus violations might be expected. However, if only a single set of lasers is used, the act of triggering the laser to measure velocity effectively gives its position (thought using the beamwidth to measure velocity really produces an *average* over the width of the beam, but a sufficiently narrow beamwidth might eventually be achievable).

## VI. INTERPRETATIONS AND CONCLUSIONS

A violation of these inequalities would actually serve to *verify* the uncertainty principle since it would indicate a definite correlation between position and momentum. In addition it would demonstrate a link between the entanglement of two particles and the correlation of position and momentum for a *single* particle. One might be tempted to refer to it as a form of *self-*entanglement, though it would be unclear just what that meant. A more direct interpretation might look at just how one were to transform, say, from a +1 to a -1. For instance, what is the mechanism by which a spin up can be turned into a spin down, or, in the present example, what is the mechanism by which position can be turned into momentum.

Take the case of position and momentum, for instance. Certainly the most accurate method for measuring position, at least classically, is to measure from a reference frame co-moving with the object being measured meaning the object is at rest in that frame. A Lorentz boost to another *non*-comoving frame provides a relative velocity between the frames and thus a measurable momentum in the second frame. Similarly one can perform a Lorentz rotation to transfer from a frame in which a particle is spin down to one in which it is spin up. This might point to a field theoretic explanation. Though scalar fields commute and seemingly prevent any superluminal signaling, even in measurement, vector fields include *anti*-commutators [11-12]. This does not normally present a problem for causality because observables are generally built out of an *even* number of spinor fields [12]. However, it is clear from experiment that, at least in the usual [3+1] dimensions, causality is *not* preserved. Rather than attempting to find a causal explanation for entanglement, it may be more useful to find a way to exploit the anti-commuting nature of vector fields to build a set of observables out of an *odd* number of spinor fields. Single-particle entanglement might be described by a self-interacting field.

As a final note, generalized work on quantum mechanical joint measurements of non-commuting observables is not a new idea [13-15]. The work of Arthurs and Kelly dates from 1965, for instance, and Cirel'son's work from 1980 offers a slightly weaker bound on (4). The presentation here offers suggestions for experiments that utilize more recent advances in technology and points to a possible field theoretic explanation for the phenomenon of entanglement.

In addition since all non-commuting observables obey an uncertainty relation and, theoretically, can be shown to obey some set of Bell-type inequalities (depending on the proposed measurement of the observables) there should be some relation between these two types of relations. In fact a set theoretic relation has been recently derived [16].

### Acknowledgements

I thank Chao-Yang Lu for several helpful comments on the first version of this paper. I also thank Steven French and Jeff Schnick for useful comments on many of the concepts contained within the paper, as well as Will Murphy and Jochen Heisenberg for serving as useful "sounding boards" on two notable occasions.